\documentclass[aip,jcp,reprint,floatfix]{revtex4-1}
\usepackage{amsmath,amsfonts,amssymb}
\usepackage{graphicx}

\usepackage{color}

\graphicspath{{./}}

\newcommand{\ca}{Ca$^{2+}$}

\begin{document}
\title{
Coarse-Grained Molecular Simulations of Allosteric Cooperativity}

\author{Prithviraj Nandigrami and John J. Portman}

\affiliation{Department of Physics, Kent State University, Kent, OH 
  44242 } 

\date{\today}

\begin{widetext}

\begin{abstract}

\noindent
Interactions between a protein and a ligand are often
accompanied by a redistribution of the population of thermally accessible
conformations.  This dynamic response of the protein's functional energy
landscape enables a protein to modulate binding affinities and control
binding sensitivity to ligand concentration.  In this paper, we
investigate the structural origins of binding affinity and allosteric
cooperativity of binding two \ca ions to each domain of calmodulin (CaM)
through simulations of a simple coarse-grained model.  In this model, the
protein's conformational transitions between open and closed
conformational ensembles are simulated explicitly and ligand binding and
unbinding is treated implicitly 
\color{black}
within the Grand Canonical Ensemble.  
\color{black}
Ligand binding
is cooperative because the binding sites are coupled through a shift in
the dominant conformational ensemble upon binding.  
The classic Monod-Wyman-Changeux model of
allostery with appropriate binding free energy to the open and closed ensembles 
accurately describes the simulated binding thermodynamics. 
The simulations predict that the two domains of CaM
have distinct binding affinity and cooperativity.  In particular,
C-terminal domain binds \ca with higher affinity and greater cooperativity
than the N-terminal domain. From a structural point of view, the affinity
of an individual binding loop depends sensitively on the loop's
structural compatibility with the ligand in the bound ensemble, as well as
the conformational flexibility of the binding site in the unbound
ensemble. 
\end{abstract}

\end{widetext}

\maketitle


\section*{Introduction}

Conformational dynamics is essential for a protein's ability to
exhibit allostery. The coupling between two distant binding sites is frequently
accomplished by a conformational change between a ``closed'' (apo) to an
``open'' (holo) conformation upon ligation.\cite{goh:04} Although the end point conformations
often give valuable insight into protein function, a detailed
description of the allosteric mechanism for a particular protein requires one to
consider a broader conformational ensemble. The landscape theory of
binding\cite{ma:99,swain:06,boehr:09} acknowledges that a folded protein is
inherently dynamic and explores the thermally accessible conformational states
in its native basin.\cite{henzler-wildman:07} This conformational ensemble
comprises the protein's ``functional landscape''.\cite{zhuravlev:10b} While only
a small subset of the states comprising the folding energy
landscape\cite{bryngelson:95}, the functional landscape determines how a protein
responds to the changes in its local environment such as ligand
interactions. Due to the heterogeneous nature of the conformational ensemble, a
ligand preferentially stabilizes some conformations more than others, causing the
protein's thermal population to redistribute to a ligated ensemble which in
general has distinct equilibrium properties.\cite{kumar:00,smock:09} The
ensemble nature of allostery accommodates a rich and diverse set of regulatory
strategies and provides a general framework to understand binding thermodynamics
and kinetics of specific proteins.\cite{hilser:12,motlagh:14} Even simple
landscapes with a small number of well defined basins separated by kinetic
barriers can have subtle binding mechanisms because they depend on ligand
interactions to short-lived transient states. Experimental progress on this
challenging kinetics problem has appeared only very
recently.\cite{daniels:15} In principle, affinities of metastable states can
also be obtained from thermodynamic binding measurements, although such analysis
may not always be practical. In this paper, we focus on the cooperative binding
of two \ca ions to the binding loops of the domains of Calmodulin (CaM) through
equilibrium coarse-grained simulations.

In this minimal model, the conformational transition between the open and closed
ensembles are simulated explicitly and the dynamic shift in population due to
ligand binding and unbinding is approximated by discrete jumps between a ligated
and unligated free energy surfaces.\cite{kenzaki:11} The protein dynamics are
governed by a native-centric potential that couples the open and closed
conformational basins while ligation is represented implicitly through
ligand mediated protein contacts. This model, developed by Takada and
co-workers, has been used to investigate the kinetic partitioning of induced fit
and conformational selection binding pathways\cite{okazaki:08} as well as
mechanical unfolding of Calmodulin in the presence of \ca.\cite{li:14} 
\color{black}
Here, we assume that the ligands bound to the protein are in equilibrium with a dilute 
solution and calculate binding thermodynamics as a function of ligand concentration.
\color{black}

The model is parameterized so that the closed basin is more stable than the open
basin in the unligated ensemble. 
\color{black}
%
\color{black} Ligands interact with all conformations in the ensemble, but the
affinity is largest for conformations within the open basin due to their high
structural compatibility with the ligand. Thus, the population shifts towards
the open ensemble with increasing ligand concentration (see \textit{SI} Fig.S1).
The simulated ensembles have significant molecular fluctuations which modulate
ligand affinities and affect the coupling between the binding sites. When
binding thermodynamics are dominated by the open and closed ensembles, this
model provides a molecular realization of the celebrated Monod-Wyman-Changeux
(MWC) model of allostery.\cite{monod:65,changeux:12} 
\color{black}
For binding a single ligand the MWC model has four states:
unligated-open, unligated-closed, ligated-open, and ligated-closed.
\color{black}
Appealing to this simple four-state model
allows us to extract binding free energies of the isolated sites in the
simulated open and closed ensembles and to calculate the free energy associated
with the cooperative coupling between the sites. The simulations connect the
conformational ensemble underlying the protein's dynamics with the MWC
phenomenological binding parameters.\cite{cui:08}

Early work on binding thermodynamics of CaM has revealed
that the affinities and cooperativities
of the N-terminal domain (nCaM) and the C-terminal domain
(cCaM) are distinct despite their structural similarity.\cite{klevit:84,wang:85,linse:91,bayley:96,masino:00}
Although some experimental data has been reanalyzed
recently\cite{stefan:08,stefan:09,lai:15}, the traditional analysis of thermodynamic binding data
has not used a dynamic landscape (or MWC) framework.\cite{linse:91,sorensen:98,newman:08} 
Nuclear Magnetic Resonance experiments\cite{barbato:92,evenas:99,evenas:01} 
and all atom molecular dynamics simulations\cite{vigil:01}
that show a dynamic equilibrium between the open and closed
conformations of CaM's domains in the absence of \ca support our approach.

\section*{Methods}

Calmodulin (CaM) is a small, 148 amino acid long protein consisting of two
topologically similar domains.  Each domain consists of four $\alpha$-helices and a
pair of EF-hand \ca-binding loops. The N-terminal domain (nCaM) has helices
labeled A -- D with binding loops I and II, and the C-terminal domain (cCaM) has helices
labeled E -- H with binding loops III and IV.  We simulate open/closed allosteric
transitions of the isolated domains of CaM using a native-centric model
implemented in the $\textit{Cafemol}$ simulation package 
\color{black}
developed by Takada and co-workers.\cite{kenzaki:11} 
\color{black}
This
model couples two energy basins, one biased to the open (pdb: 1cll\cite{chattopadhyaya:92}) reference
structure and the other biased to the closed (pdb: 1cfd\cite{kuboniwa:95}) reference structure.  The
energy of a conformation, specified by the $N$ position vectors of the
C-$\alpha$ atoms of the protein backbone,
$\mathbf{R} = \{\mathbf{r}_1, \cdots \mathbf{r}_N \}$, is given by
%
\begin{align}
  V({\mathbf R}) = 
  &\left( V_\mathrm{o}({\mathbf R})+V_\mathrm{c}({\mathbf R})+\Delta V\right)/2  \\
  &-\sqrt {\left(V_\mathrm{o}({\mathbf R})-V_\mathrm{c}({\mathbf R})-\Delta V\right)^2/4 + \Delta ^2} \, ,\nonumber
\end{align}
%
where $V_\mathrm{o}({\mathbf R})$ is the single basin potential defined by the
open structure and $V_\mathrm{c}({\mathbf R})$ is the single basin potential
defined by the closed structure.  The interpolation parameters, $\Delta$ and
$\Delta V$, control the barrier height and the relative stability of the two
basins. Parameters defining the single energy basins are set to their default
values with uniform contact strength.  
The simulation temperature
is set below the folding transition temperature of each of the four
conformations.  Specifically, the simulation temperature is set to
$T_\mathrm{{sim}} = 0.8T_\mathrm{F}^{\star}$ where
$T_\mathrm{F}^{\star} = 329 ^{\circ}\mathrm{K}$ is the folding transition
temperature corresponding to the closed (apo) state of nCaM, the lowest
transition temperature among the open and closed states of nCaM and cCaM.
\color{black}
Equilibrium trajectories of length $10^{8}$ steps are simulated using Langevin
dynamics with a friction coefficient of $\gamma = 0.25$ and a timestep of
$\Delta t = 0.2$ (in coarse-grained units).\cite{okazaki:08}

\color{black}
Calcium binding to the two EF-hand loops of each domain of CaM is modeled
implicitly by adding a potential defined from the ligand-mediated contacts in
the EF-hand loops of the open (holo) conformation
\begin{equation}\label{eq:binding_energy}
  V_{\mathrm{bind}} = 
  - \sum_{i,j}
   c_{\mathrm{lig}} \epsilon_{\mathrm{go}} \exp \left[ - \frac {\left(r_{ij} - r_{ij}^0\right)^2}{2 \sigma_{ij}^2}\right].
\end{equation}
Here, the sum is over pairs of residues that are each within 4.5 $\AA \,$ of a
\ca ion and closer than 10.0 $\AA \,$ in the open (holo) conformation.  The
binding energy parameters $c_\mathrm{lig}$, $\epsilon_\mathrm{go}$, and $\sigma$
are taken to be the same for each ligand-mediated contact for simplicity.

Binding cooperativity is influenced by the
relative stability of the unligated open and closed states determined by
$\Delta V$ and the binding free energy determined by $V_\mathrm{bind}$.  
In principle, these parameters can be adjusted to match measured binding
properties.  In the absence of clear measured constraints, we choose
parameters so that the relative stability between the open and closed states are
the same for each domain.

The transition barrier height is determined by $\Delta$ which is set to
$14.0 \, \mathrm{kcal/mol}$ for nCaM and $17.5 \, \mathrm{kcal/mol}$ for cCaM.
Adjusting $\Delta V = 5.0 \,\mathrm{kcal/mol}$ for nCaM and
$\Delta V = 4.75 \,\mathrm{kcal/mol}$ for cCaM while keeping other parameters
fixed gives an energy difference between the unligated open and closed states,
$\epsilon = 4 \, k_{\mathrm{B}} T$ for both domains.  
%
%
\color{black}
Experimentally, the folding temperatures of the N-terminal and C-terminal domains
in intact CaM are approximately $328^\circ \mathrm{K}$ and
$315^\circ \mathrm{K}$, respectively.\cite{rabl:02}
Connecting to the domain opening kinetics in the intact protein,
our simulation temperature corresponds to approximately $310^\circ \mathrm{K}$
which is 95\% of nCaM's simulated folding temperature, and 98\% of
cCaM's simulated folding temperature.  
\color{black}
For the results reported in this paper, the binding energy parameters 
are set to
$\epsilon_\mathrm{go}=0.3$ (default value in $\textit{Cafemol}$),
$c_{\mathrm{lig}}=2.5$ and $\sigma_{ij}=(0.1) r_{ij}^0$ where $r_{ij}^0$ is the
corresponding separation distance in the open (holo) reference conformation.
%
%
\color{black}
We have performed additional simulations to explore the dependence
of binding thermodynamics on the ligand-mediated
contact strength and interaction range.
%
\color{black}
At higher values of $c_\mathrm{lig}$ and $\sigma_{ij}$, 
the affinities of ligand binding to individual loops increase.
Nevertheless, the slope of the titration curve at the midpoint of the
transition (a measure of binding cooperativity) remains the same (\textit{SI} Fig.S2).

\color{black}
The simulated conformational ensembles are characterized structurally in terms
of local and global order parameters based on the contacts formed in each
sampled conformation.  The set of native contacts in the open and closed
conformations are separated into three groups: those that occur exclusively in
either the open or the closed native structures, and those that are common to
both states.  A native contact in a given conformation is considered to be
formed provided the distance between the two residues is closer than 1.2 times
the corresponding distance in the native conformation.  Local order parameters
$q_\mathrm{open}(i)$ and $q_\mathrm{closed}(i)$ are defined as the fraction of
native contacts involving the $i^{th}$ residue that occur exclusively in the
open and closed native structures, respectively.  Overall native similarity is
monitored by corresponding global order parameters,
$Q_\mathrm{open} = \langle q_\mathrm{open}(i) \rangle$ and
$Q_\mathrm{closed} = \langle q_\mathrm{closed}(i) \rangle$, where the average is
taken over the residues of the protein.  We identify metastable conformational
basins from minima in the free energy computed through the population
histogram parameterized by $Q_\mathrm{open}$ and $Q_\mathrm{closed}$.

\color{black}
Ligand binding/unbinding events coupled with a conformational
change of the protein is modeled within the Grand Canonical Ensemble. 
\color{black}
Throughout the protein's conformational transitions, 
the ligation
state of each loop is determined stochastically through a Monte Carlo step
attempted every 1000 steps in the Langevin trajectory.  If the loop is
unligated, a ligand is introduced to the binding loop
($V\rightarrow V + V_\mathrm{bind}$) with probability
\begin{equation}
  \label{eq:metropolis1}
  P_{0\rightarrow 1} = \min [ 1, \exp \left[-(V_{\mathrm{bind}}-\mu)/k_\mathrm{B}T \right]].
\end{equation}
If the loop is ligated, the ligand dissociates from the binding loop 
($V + V_\mathrm{bind} \rightarrow V$) with probability
\begin{equation}
  \label{eq:metropolis2}
  P_{1\rightarrow 0}  = \min [ 1, \exp \left[(V_{\mathrm{bind}}-\mu)/k_\mathrm{B}T \right]].
\end{equation}
Here, $\mu$ is the chemical potential of a bound ligand.  At equilibrium, $\mu$
equals the chemical potential of the ligand in solution, 
\begin{equation}
  \label{eq:chem_potential}
  \mu = k_{\mathrm{B}} T \ln \left( \frac{c}{c_0}\right)  + \mu_0 \, ,
\end{equation}
where $c$ is the ligand concentration, and $c_0$ and $\mu_0$ are the
reference concentration and reference chemical potential, respectively.  
\color{black}
To compute binding curves, a series of simulations are preformed,
each at a different value of the ligand chemical potential. 
\color{black}
These simulated titration curves 
are reported as
function of the chemical potential, or equivalently, in terms of the relative
ligand concentration defined through
$\mu/k_\mathrm{B}T = \ln \left(c/\bar{c}_0\right)$ where
$\bar{c}_0 = c_0\exp(-\mu_0/k_\mathrm{B}T)$.

\color{black} 
This approach with Monte Carlo acceptance rates given in Eq.~\ref{eq:metropolis1} and
Eq.~\ref{eq:metropolis2} is oriented
towards binding thermodynamics from the outset.  Takada and co-workers present a
different choice motivated by ligand binding kinetics.\cite{okazaki:08}  Instead of
introducing a chemical potential, ligand concentration enters their model through
a variable binding attempt rate, while the attempt rate of unbinding is fixed.  Binding
titration curves can also be calculated in this model, but as a function of the binding
attempt rate rather than the concentration directly.\cite{li:14}
\color{black}

\section*{Binding a single ligand}
\begin{figure*}
  \begin{center}
    \includegraphics[]{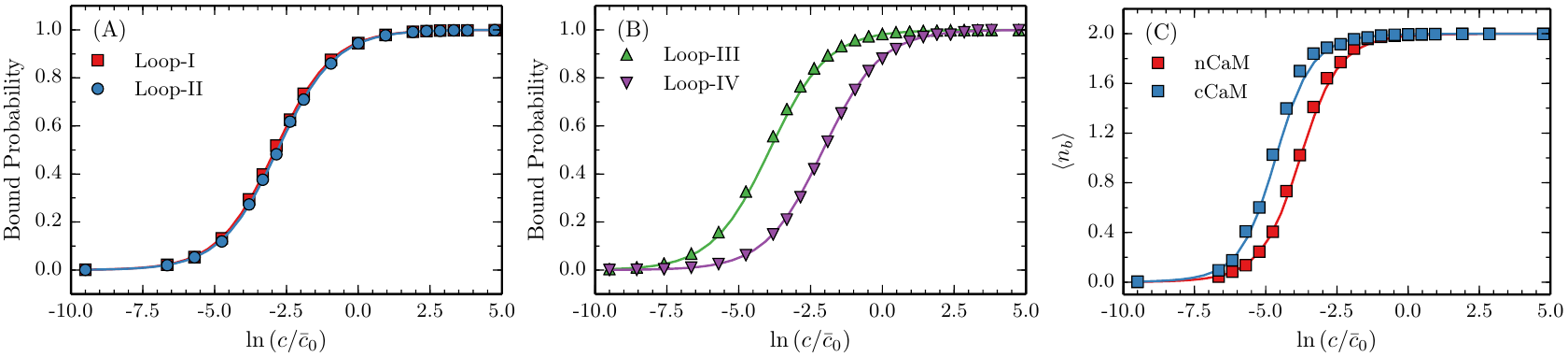}
    \caption{Simulated binding curves for the individual loops of (A) nCaM and
      (B) cCaM.  Lines are fits to the two state MWC model given by
      Eq.\ref{eq:bound_prob_one_lig}.  (C) Simulated mean number of bound ligands
      occupancy of binding sites with two ligands for nCaM (blue) and cCaM (red)
      as a function of ligand concentration.  The solid lines plot
      $\langle n_\mathrm{b}(\mu) \rangle = p^\mathrm{A0}_\mathrm{b}(\mu) +
      p^\mathrm{0B}_\mathrm{b}(\mu) + 2 p^\mathrm{AB}_\mathrm{b}(\mu) $
      with probabilities given from the MWC model evaluated with the binding
      parameters found from fits of binding to each individual loops. }
    \label{fig:ncam_ccam_affinity}
  \end{center}
\end{figure*}

We first consider \ca binding exclusively to each individual loop 
\color{black}%
by simulating the conformational change of the entire domain while permitting
binding only to a single site.  As shown in Fig.\ref{fig:ncam_ccam_affinity} (A)
and Fig.\ref{fig:ncam_ccam_affinity} (B), 
\color{black} 
the bound population as
a function of ligand concentration, $p_\mathrm{b}(c)$, follows a typical
sigmoidal profile connecting a fully unbound population at low concentration and
a fully bound population at high concentration. The overall binding strength of
the individual loops is reflected in the dissociation constant, $K_\mathrm{d}$,
shown in Table.\ref{tab:single_loop_data}. Binding affinities of nCaM's loops
are nearly the same, whereas the affinities of cCaM's loops are significantly
different, with $\mathrm{K_{d}^{(IV)}} \approx 7 \,\mathrm{K_{d}^{(III)}}$.
Comparing the binding strength of CaM's loops, our simulations predict that
$\mathrm{K_{d}^{(III)}} < \mathrm{K_{d}^{(I)}} \approx \mathrm{K_{d}^{(II)}} <
\mathrm{K_{d}^{(IV)}}$.

\begin{table}

\color{black}
\caption{Number of ligand-mediated contacts, dissociation constants,
and binding free energies for the loops of CaM.}
\color{black}
\begin{ruledtabular}
\begin{tabular}{lcccc}
& $N_\mathrm{con}$
& $K_\mathrm{d}/\bar{c}_0$
& $\epsilon_\mathrm{c}$\footnotemark[1] & $\epsilon_\mathrm{o}$\footnotemark[1]\\
\hline
loop I & 5 & 0.054 & -1.3 & -3.1 \\
loop II & 5 & 0.062 & -1.3 & -2.9\\
loop III & 8 & 0.018 & -1.3 & -4.1\\
loop IV & 5 & 0.13 & -0.5 & -3.0\\
\hline
\end{tabular}
\label{tab:single_loop_data}
\footnotetext{in kcal/mol}
\end{ruledtabular}
\end{table}

It is reasonable to expect that binding affinities from a uniform native-centric
model correlate with the number of ligand mediated contacts, $N_\mathrm{con}$.  
While loop III
does indeed have the most contacts and the greatest binding affinity, accounting for 
reduced affinity of loop IV compared to the loops of nCaM, each with the same
number of contacts, requires more careful explanation.  
Such subtlety is not surprising because binding strength is 
sensitive to a protein's conformational flexibility that modulates ligand interactions 
in both the open and closed ensembles. 
\begin{figure}
  \begin{center}
    \includegraphics[]{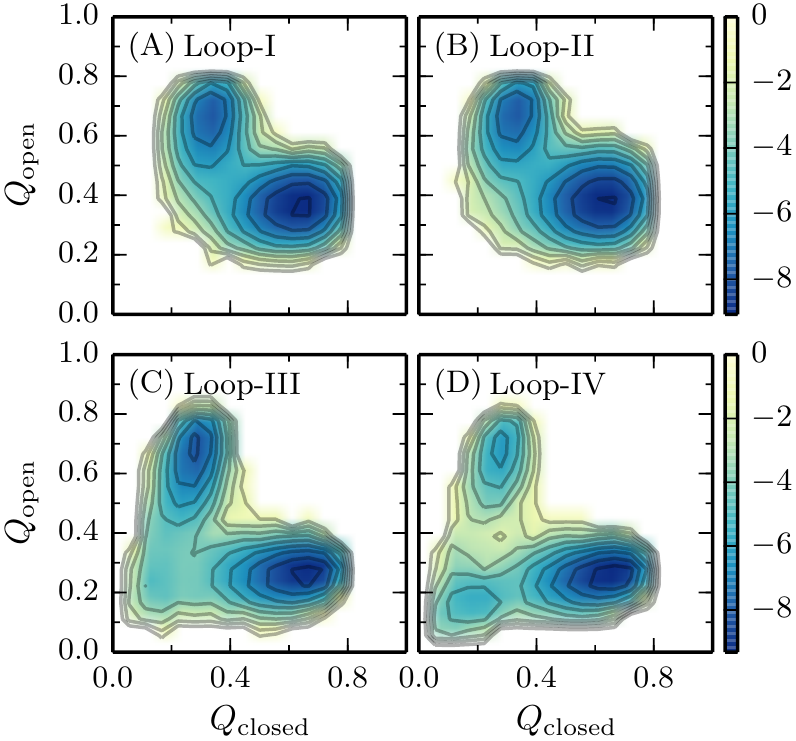}
    \caption{Simulated free energy as a function of
      $Q_{\mathrm{closed}}$ and $Q_{\mathrm{open}}$
      for binding loops at ligand concentration
      $\mathrm{c} = K_\mathrm{d}$
      for nCaM (A,B) and  cCaM (C,D).
      The open state ensemble are conformations with
      $0.18 \le Q_{\mathrm{closed}} \le 0.35$
      and $0.55 \le Q_{\mathrm{open}} \le 0.75$.}
    \label{fig:free_energy_contour}
  \end{center}
\end{figure}
The MWC model provides insight into the affinities for the individual binding
loops.  Using notations in Ref.\onlinecite{marzen:13}, the bound population in the MWC
model can be expressed as
\begin{equation}\label{eq:bound_prob_one_lig}
  p_{\mathrm{b}}(\mu) = \left( e^{-\beta(\epsilon_\mathrm{{c}} - \mu)}
    + e^{-\beta(\epsilon + \epsilon_\mathrm{{o}} - \mu)}\right)/Z_1
\end{equation}
where 
\begin{equation}\label{eq:partition_func_one_lig}
  Z_1 = 1 + e^{-\beta(\epsilon_\mathrm{{c}} - \mu)} 
  + e^{-\beta \epsilon} \left( 1 + e^{-\beta(\epsilon_\mathrm{{o}} - \mu)}\right) 
\end{equation}
is the single ligand partition function. Here,
$\epsilon_\mathrm{{c}}$ and $\epsilon_\mathrm{{o}}$ denote the binding free
energies of the ligand to the closed and open ensemble, $\epsilon$ is the
difference in stability between the unbound closed and open ensemble, and $\mu$
is the ligand chemical potential. 

\begin{figure}
  \begin{center}
    \includegraphics[]{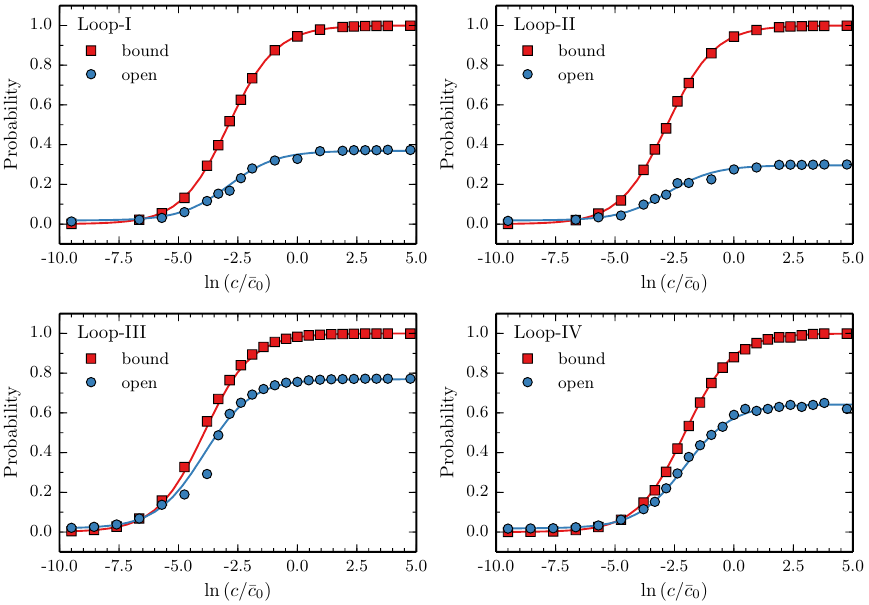}
    \caption{Simultaneous fits of simulation data for a single ligand 
      to $p_\mathrm{b}(\mu)$ and $p_\mathrm{o}(\mu)$
      for individual binding loops.
      Solid curves are plots of $p_\mathrm{b}(\mu)$ and $p_\mathrm{o}(\mu)$
      with $\epsilon_\mathrm{c}$ and $\epsilon_\mathrm{o}$ determined by a 
      simultaneous fit to the simulation data (shown as points).}
    \label{fig:prob_bound_open}
  \end{center}
\end{figure}

The coupling parameters of the
Hamiltonian fix 
$\epsilon = 4k_\mathrm{B}T$ in the simulation,
leaving the binding parameters,
$\epsilon_\mathrm{c}$ and $\epsilon_\mathrm{o}$, to be determined from the
simulated titration curves.  
These two parameters are under-determined by a fit
to the bound state probability alone. The population of the open ensemble
(regardless of ligation state) provides an additional constraint for parameters
in the model. The open state configurations are identified in the simulations
by global order parameters that measure the similarity to the open and closed
native structures (see Fig.\ref{fig:free_energy_contour}). 
\color{black} At high ligand concentration, the bound population
saturates to unity.  Since ligands bind to both the open and closed
conformations, the limiting value of the open population, $p_o(\mu)$, is the
fraction of ligated protein in an open conformation.
As shown in Fig.\ref{fig:prob_bound_open},
the simulated open population of each loop saturates to a different 
limiting value determined by the relative stability of binding to the open and closed
ensembles. 
\color{black}

In the MWC model, the open state population
\begin{equation}\label{eq:open_prob_one_lig}
  p_\mathrm{o}(\mu) = 
    e^{-\beta \epsilon}\left(1 + e^{-\beta(\epsilon_\mathrm{{o}} - \mu)}\right)/Z_1
\end{equation}
has a limiting value, $p_{o}(\beta \mu \gg 1) \sim [1 + e^{\beta(\epsilon
  +\Delta \epsilon)}]^{-1}$, that depends on $\Delta \epsilon = \epsilon_{o}
-\epsilon_{c}$.  Thus, $p_\mathrm{o}(\mu)$ and $p_\mathrm{b}(\mu)$ are
independent constraints that can be used to determine reliable model parameters
for the open and closed binding free energies, $\epsilon_\mathrm{o}$ and
$\epsilon_\mathrm{c}$. 

The binding free energies determined by a simultaneous fit to
Eq.\ref{eq:bound_prob_one_lig} and Eq.\ref{eq:open_prob_one_lig} are shown in
Table.\ref{tab:single_loop_data}. As expected, the dissociation constants
depend on both $\epsilon_\mathrm{c}$ and $\epsilon_\mathrm{o}$. The values
of $\epsilon_\mathrm{o}$ tracks the number of ligand mediated contacts in each
loop, with loop III being the most stable, while the other loops have similar
stability.  The values of $\epsilon_\mathrm{c}$ is more subtle. Although loop
III has more contacts than loop I and loop II, they all have the same
$\epsilon_\mathrm{c}$. Thus, the relatively high affinity of loop III can be
attributed to its greater stabilization upon binding to the open state.  The
relatively low affinity of loop IV, in contrast, is explained by the smaller
binding stabilization to the closed state.  

Although experiments suggest that CaM's binding loops have 
heterogeneous affinities, assigning a binding strength to each loop is 
challenging because techniques to distinguish site-specific binding 
tend to alter the stability of the open and closed states.\cite{gifford:07}
Early studies which isolate the binding
properties of loops III and IV of cCaM through site-directed mutagenesis
indicate that loop IV has a higher propensity of \ca-binding than loop III
.\cite{evenas:98,malmendal:99} A similar approach indicates the
affinity of the nCaM's loops are comparable, with loop I reported to have only
1.5 times higher affinity than the affinity of loop II.\cite{beccia:15} 
On the other hand, isolating the loops by grafting them to a scaffold
suggests a different order, with 
loop I having the highest affinity and 
loop III binding \ca more tightly than loop IV.\cite{ye:05} 
Validation of our simulation results would
benefit from experimental clarification of the relative 
binding affinities of the loops of CaM.

The effective 
binding free energies represent average properties over an ensemble that
may include a broader range of conformations than those near the open and closed state minima. 
As shown in Fig.\ref{fig:free_energy_contour}, the
conformational ensemble
for binding to nCaM's loops are two state while the binding to cCaM's loops
includes contribution from a partially unfolded basin as well.
\color{black}%
The appearance of an unfolded intermediate in the domain opening transition
of \ca-free cCaM was first reported by Chen and Hummer.\cite{chen:07}
\color{black} 
Distinct 
ensembles for nCaM and cCaM are consistent with the simulated transition
mechanisms of the domains in the absence of \ca (submitted). 
Although the four-state description of MWC is an approximation
(especially for cCaM), it's use is validated by the accurate description of the
populations of different ligation states for simulations of two ligand binding
discussed in the next section.

\section*{Binding two ligands}
We turn now to simulations in which both binding sites are accessible to the ligands.
The mean number of bound ligands as a function of concentration,
shown in 
\color{black} %
Fig.\ref{fig:ncam_ccam_affinity} (C), 
\color{black}
indicates that the effective dissociation
constant of nCaM, $K_\mathrm{d}(\mathrm{nCaM})/\bar{c}_0 = 2.2 \times 10^{-2}$,
is roughly three times larger than the dissociation constant of cCaM,
$K_\mathrm{d}(\mathrm{cCaM})/\bar{c}_0 = 8.8 \times 10^{-3}$.  For both domains,
the mid-point concentration for binding two ligands is smaller than the dissociation
constants for the individual binding sites in the domain.  The finding that cCaM
has greater overall binding affinity than nCaM agrees qualitatively with
experiments.\cite{bayley:96,stigler:12}
Additionally, the estimated value of $K_\mathrm{d}(\mathrm{nCaM})$ is within the 
experimentally reported range of approximately 6 -- 10 times
$K_\mathrm{d}(\mathrm{cCaM})$.\cite{klevit:84,wang:85,linse:91}

Binding curves calculated within the MWC model are also shown in
\color{black}%
Fig.\ref{fig:ncam_ccam_affinity} (C).
\color{black}  
Denoting the two binding sites in the domains as
site $A$ (loop I or loop III) and site $B$ (loop II or loop IV), we calculate
$\langle n_\mathrm{b}(\mu) \rangle = p^\mathrm{A0}_\mathrm{b}(\mu) +
p^\mathrm{0B}_\mathrm{b}(\mu) + 2 p^\mathrm{AB}_\mathrm{b}(\mu) $
where 
\begin{align}
  p_\mathrm{{b}}^{\mathrm{A0}}(\mu) 
  &= ( 
  e^{-\beta(\epsilon_\mathrm{{c}}^{\mathrm{A}}- \mu) }
  + e^{-\beta (\epsilon + \epsilon_\mathrm{{o}}^{\mathrm{A}} - \mu)} 
  )/Z_2      \label{eq:prob_only_1_bound} \\
  p_\mathrm{{b}}^{\mathrm{0B}}(\mu) 
  &= ( 
  e^{-\beta(\epsilon_\mathrm{{c}}^{\mathrm{B}} - \mu)}
  + e^{-\beta (\epsilon + \epsilon_\mathrm{{o}}^{\mathrm{B}} - \mu)} 
  )/Z_2\label{eq:prob_only_2_bound} \\
  p_\mathrm{{b}}^{\mathrm{AB}}(\mu) 
  &= ( 
  e^{-\beta(\epsilon_\mathrm{{c}}^{\mathrm{A}} + \epsilon_\mathrm{{c}}^{\mathrm{B}} - 2\mu)} 
  + e^{-\beta (\epsilon + \epsilon_\mathrm{{o}}^{\mathrm{A}} + \epsilon_\mathrm{{o}}^{\mathrm{B}} -2\mu)}
  )/Z_2\label{eq:prob_both_bound}
  \end{align}
are the probabilities for the
conformational ensemble with site A occupied and site B empty, site B
occupied and site A empty, and both binding sites simultaneously occupied,
respectively. Here, 
$Z_2 = Z_\mathrm{c} + e^{-\beta \epsilon} Z_\mathrm{o}$ denotes the two ligand partition function
with 
\begin{equation}
Z_i =
1 + e^{-\beta(\epsilon_i^{\mathrm{A}} - \mu)} + e^{-\beta(\epsilon_i^{\mathrm{B}} - \mu)},
\quad
i = \mathrm{(o,c)}.
\end{equation}
where, for example, 
$\epsilon_\mathrm{{c}}^\mathrm{A}$ and $\epsilon_\mathrm{{o}}^\mathrm{A}$
denotes the binding free energies to loop A (described
in the previous section).  The agreement between the MWC model and simulated
binding curves is excellent, indicating that the binding cooperativity of the
simulations is well characterized by the MWC Model.

\begin{figure}
  \centering
  \includegraphics[]{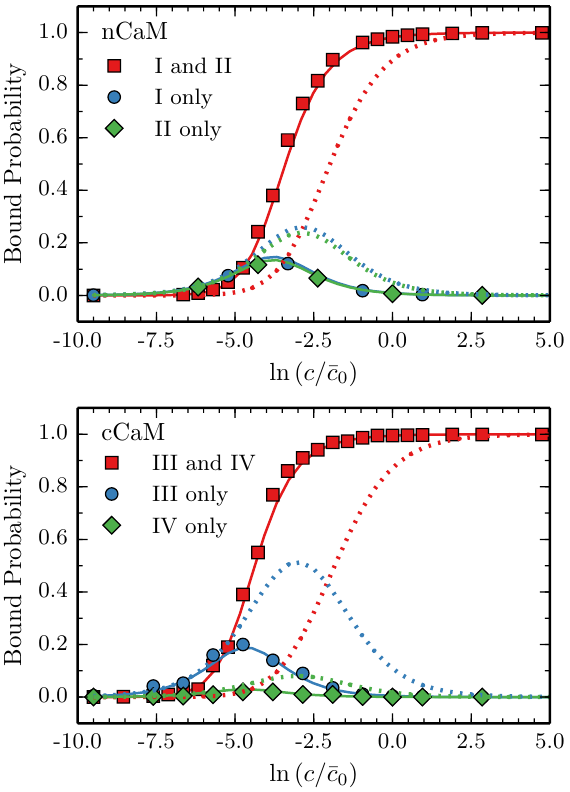}
  \caption{Populations of ligation states $p_\mathrm{{b}}^{\mathrm{A0}}(\mu)$
    (blue), $p_\mathrm{{b}}^{\mathrm{0B}}(\mu)$ (green), and
    $p_\mathrm{{b}}^{\mathrm{AB}}(\mu)$ (red) plotted as a function of \ca
    concentration for nCaM (top) and cCaM (bottom).  Simulation data shown as
    points. Solid curves plot
    Eq.(\ref{eq:prob_only_1_bound}--\ref{eq:prob_both_bound}) from the MWC
    model.  Dotted curves show plots of the non-cooperative induced fit model of
    binding to independent sites described by the partition function given in
    Eq.\ref{eq:z_indep_binding}.  Note some data points are skipped for
    clarity.
    \label{fig:ncam_ccam_mwc_compare}
}
\end{figure}

The MWC model also quantitatively captures the simulated populations of
individual ligation states as shown in 
Fig.\ref{fig:ncam_ccam_mwc_compare}.
Starting at low concentration, the growth of the singly ligated and the fully
ligated states are concomitant.  The fully loaded protein becomes increasingly
stable, thereby reducing the singly ligated populations after certain
threshold.  The probability of exclusive binding to either loop of nCaM is
equal, attaining a maximum population of 15\% each. In contrast, virtually all
of binding of the first ligand in cCaM occurs in loop III, reaching a maximum
population of 20\%.  The near complete suppression of \ca ligation exclusively to
loop IV is due to its small relative binding affinity.

\section*{Binding cooperativity}
In cooperative binding, enhanced recruitment of a second ligand suppresses the
population of singly ligated proteins thereby sharpening the binding
curve. Within the assumptions of the MWC model, the shape of the binding curve
is determined by two mechanisms. First, the greater stabilization of the open
conformation over the closed conformation upon binding 
makes even binding a single ligand more sensitive to changes in concentration.  
The second  source of cooperativity is the
allosteric coupling provided by the assertion that the binding sites are either
both open or both closed depending on the conformational state of the entire
protein.

Comparing the simulated binding curves to those produced from a model that
neglects both of these cooperative assumptions gives a qualitative sense of the
simulated binding cooperativity.  We consider the binding probabilities
calculated according to the partition function for induced fit binding to
independent binding sites. These binding probabilities, shown in
Fig.\ref{fig:ncam_ccam_mwc_compare}, are calculated according to the partition
function $Z_\mathrm{IF} = Z_\mathrm{A}Z_\mathrm{B}$ where
\begin{equation}
  \label{eq:z_indep_binding}
  Z_\alpha = 
  \left( 1 + e^{-\beta (\epsilon_\mathrm{o}^\mathrm{\alpha}-\mu)} 
    + e^{-\beta (\epsilon_\mathrm{c}^\mathrm{\alpha}-\mu)}\right),
\quad 
\alpha = (A,B).
\end{equation} 
Compared to the simulations, the populations $p_\mathrm{b}^{\mathrm{A0}}(\mu)$
and $p_\mathrm{b}^{\mathrm{0B}}(\mu)$ for independent binding to loops I -- III
initiate growth at smaller concentrations relative the the mid-point of
$p_\mathrm{b}^{\mathrm{AB}}(\mu)$, and achieve a greater maximum.  Exclusive
binding to loop IV does not develop significant population even when the loops
are independent. 
Comparing the two domains, the singly ligated states are suppressed more in
cCaM's loop III than either of nCaM's loops.  
Furthermore, the binding curve sharpens more in cCaM than nCaM. 
These comparisons show
that the simulated binding is indeed cooperative with cCaM having greater
binding cooperativity than nCaM in qualitative agreement to
experiments.\cite{linse:91,masino:00}

\begin{figure}
  \centering
  \includegraphics[width=3.in]{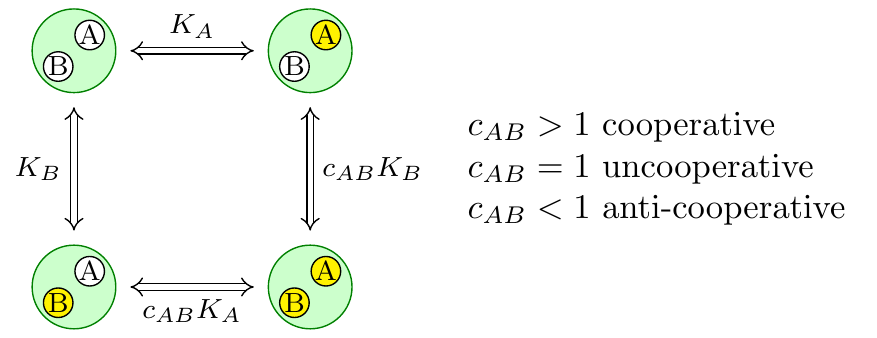}
  \caption{Thermodynamic cycle for binding two ligands.  
  \label{fig:coop_illus}
}
\end{figure}

The strength of the binding cooperativity for each domain can be determined
quantitatively by considering the thermodynamic cycle shown in
Fig.\ref{fig:coop_illus}.    
A \ca ion can bind to either loop A or loop B of
the unligated protein with equilibrium constant $K_\mathrm{A}$ and
$K_\mathrm{B}$, respectively.  The change in stability upon a \ca ion binding to
site B when site A is occupied, for example, can be expressed as
$c_\mathrm{AB}K_\mathrm{B}$ where $c_\mathrm{AB}$ represents the additional
stability associated with the presence of a previously bound ligand to site A.
A similar argument gives the equilibrium constant $c_\mathrm{AB}K_\mathrm{A}$
representing the change in stability when a \ca ion binds to site A if site B is
already occupied.  The overall equilibrium constant of the fully ligated protein
relative to the unligated protein is given by
$K_2 = c_\mathrm{AB}K_\mathrm{A}K_\mathrm{B}$ which corresponds to the binding free
energy, $\Delta F = -k_\mathrm{B}T\ln K_2$. 
The free energy associated with allosteric interactions between the ligands 
is therefore given by 
$\Delta F_{AB} = -k_\mathrm{B}T\ln c_\mathrm{AB}$.

In order to calculate $c_\mathrm{AB}$ for the simulated transitions, we 
express the equilibrium constants of the thermodynamic cycle in terms of populations
of ligation states
\begin{equation}
  \label{eq:coop_KA_KB}
  K_\mathrm{A} = p_\mathrm{b}^{\mathrm{A0}}(\mu)/[p_\mathrm{ub}(\mu) \, c],
  \quad K_\mathrm{B} = p_\mathrm{b}^{\mathrm{0B}}(\mu)/[p_\mathrm{ub}(\mu) \, c] 
\end{equation}
and
\begin{equation}
  \label{eq:coop_KAB}
  c_\mathrm{AB}K_\mathrm{A} K_\mathrm{B} = p_\mathrm{b}^{\mathrm{AB}}(\mu)/[p_\mathrm{ub}(\mu) \, c^{2}],
\end{equation}
where $p_\mathrm{ub}(\mu)$ denotes the unbound population and 
$c$ stands for the ligand concentration.
Solving for $c_\mathrm{AB}$ gives 
\begin{equation}
  \label{eq:cab}
  c_\mathrm{AB} 
  = \frac{p_\mathrm{b}^\mathrm{AB}(\mu)p_\mathrm{ub}(\mu)}{p_\mathrm{b}^\mathrm{A0}(\mu) p_\mathrm{b}^{0B}(\mu)}
\end{equation}
in terms of the population of ligated states. 
The value of $c_\mathrm{AB} = \exp(-\beta \Delta F_\mathrm{AB})$ reflects the
degree of cooperativity of the transition.  When the sites are independent,
$p_\mathrm{b}^{\mathrm{AB}}(\mu) =
p_\mathrm{b}^{\mathrm{A}}(\mu) p_\mathrm{b}^{\mathrm{B}}(\mu)/p_\mathrm{ub}(\mu)$,
so that $c_\mathrm{AB} = 1$ as expected for uncoupled sites.  

\begin{table}[htb]
\caption{
Simulated microscopic and macroscopic equilibrium constants
}
\begin{ruledtabular}
\begin{tabular}{lccccc}
& $K_\mathrm{A}$\footnotemark[1]
& $K_\mathrm{B}$\footnotemark[1]
& $c_\mathrm{AB}$ 
& $K_1$\footnotemark[1]
& $K_2$\footnotemark[2]\\
\hline
nCaM & 15.2 & 14.1 & 6.8 & 29.1 & $1.5\times 10^3$ \\
cCaM & 40.0 & 6.7 & 29.6 & 46.7 & $8.0\times 10^3$\\
\hline
\end{tabular}
\label{tab:equil_const}
\footnotetext{in units of $\bar{c}^{-1}_0$} 
\footnotetext{in units of $\bar{c}^{-2}_0$}
\end{ruledtabular}
\end{table}

Although the right hand side of Eq.\ref{eq:cab} can be evaluated directly
from simulated populations, it is convenient to take advantage of the parameterization provided by 
the MWC model since it accurately describes the simulated equilibrium populations. Using
Eq.(\ref{eq:prob_only_1_bound}--\ref{eq:prob_both_bound}) with
$p_\mathrm{ub}(\mu) = (1 + e^{-\beta \epsilon})/Z_2$ leads to
\begin{equation}
\label{eq:coop_param_mwc}
c_{\mathrm{AB}} = 
\frac{
  \left[ 1 + \exp(- \beta \epsilon)\right]
    \left[ 1 + \exp(-\beta (\epsilon + \Delta\epsilon^{\mathrm{A}} + \Delta\epsilon^{\mathrm{B}}))\right]
  }
  {
  \left[ 1 + \exp(-\beta ( \epsilon +  \Delta \epsilon^{\mathrm{A}}))\right]
  \left[ 1 + \exp(-\beta ( \epsilon +\Delta \epsilon^{\mathrm{B}}))\right]
   },
\end{equation}
with
$\Delta \epsilon^{\mathrm{A}} = \epsilon_\mathrm{o}^\mathrm{A} -
\epsilon_\mathrm{c}^\mathrm{A}$
and
$\Delta \epsilon^{\mathrm{B}} = \epsilon_\mathrm{o}^\mathrm{B} -
\epsilon_\mathrm{c}^\mathrm{B}$.
The computed equilibrium constants, shown in 
Table.\ref{tab:equil_const},
indicate that \ca-binding to cCaM (with $c_\mathrm{AB} \approx 29.6$) is more
cooperative than \ca-binding to nCaM (with $c_\mathrm{AB}\approx 6.8$) in
qualitative agreement with
experiment.\cite{linse:91,masino:00}  The cooperative free
energy is estimated to be $\Delta F_\mathrm{AB} \approx -3.4 \: k_\mathrm{B} T$
for cCaM and $\Delta F_\mathrm{AB} \approx -1.9 \: k_\mathrm{B} T$ for nCaM.
The cooperative free energy for cCaM is 1.8 times that of nCaM in agreement
with the experimental measured range of relative free energies of 1.2 --
3 reported in Ref.\onlinecite{linse:91}
and Ref.\onlinecite{waltersson:93}.

Binding thermodynamics determined from experiments that can not distinguish
between binding to individual sites are often reported through the macroscopic
equilibrium constants $K_{1} = K_{A} + K_{B}$ and
$K_{2} =
c_{AB}K_{A}K_{B}$.\cite{adair:25,waltersson:93,sorensen:98,newman:08}
The macroscopic equilibrium constants describing the simulated binding
thermodynamics are shown in Table.\ref{tab:equil_const}. The value of $K_1$ for
cCaM is greater than $K_1$ for nCaM by a factor of 1.5 in agreement with the
experimentally reported range of 1.2 -- 2.2.\cite{linse:91,waltersson:93}  The
free energy of binding two \ca ions can be estimated from the macroscopic
binding constants summarized in Table.\ref{tab:equil_const}, 
$\Delta \, G_\mathrm{tot} = -k_\mathrm{B}T \, \log(K_1K_2)$. The simulated
relative values of $\Delta \, G_\mathrm{tot}$ for cCaM is approximately 1.5 times
the value of $\Delta \, G_\mathrm{tot}$ for nCaM, which is in agreement with experimentally reported value
of approximately 1.1 -- 1.3.\cite{linse:91,jiang:10} Taken together, the simulated values of
the macroscopic binding constants for CaM are in qualitative agreement with
those reported from experiments.

\section*{Molecular description of ligand binding}
\begin{figure}
  \centering
  \includegraphics[]{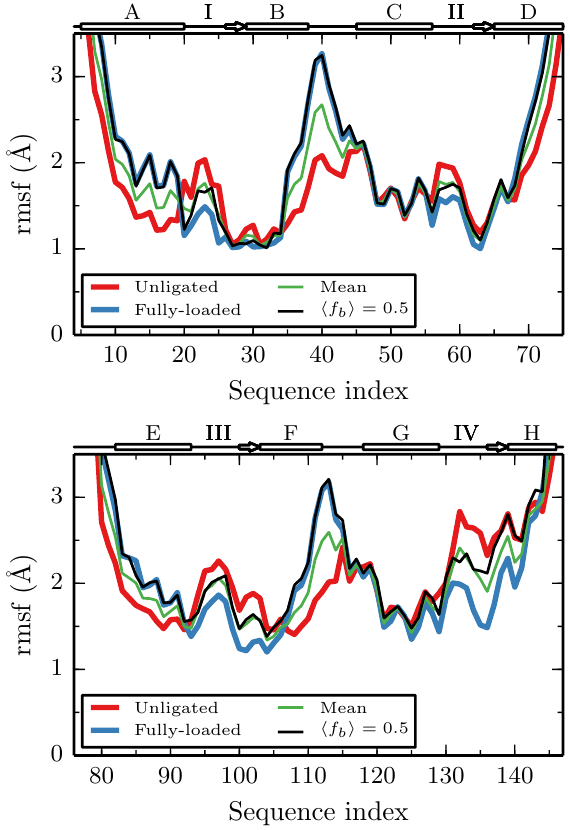}
  \caption{Simulated root mean square fluctuations (rmsf) for each residue 
    for nCaM (top) and cCaM (bottom)
    calculated at different ligand concentrations: high ligand
    concentration gives the fully saturated ensemble (blue curve), 
    low ligand concentration gives the unligated ensemble (red curve), 
    and at $K_\mathrm{d}$ (black curve). 
    The rmsf curves are calculated for each ensemble after aligning to
    the open native conformation.  (Aligning to the closed conformation give
    similar curves.)  Also shown is the reference fluctuations given in 
    Eq.\ref{eq:rms avg} (green curve).}
  \label{fig:ncam_ccam rms variation}
\end{figure}

The simulations offer a detailed molecular description of \ca binding as well as
insight into the conformational ensembles underlying the binding free energies,
$\epsilon_\mathrm{c}$ and $\epsilon_\mathrm{o}$.
Fig.\ref{fig:ncam_ccam rms variation} 
shows the root mean square fluctuations (rmsf) of each residue for
the unligated (closed) ensemble at low ligand concentration and the fully
saturated (open) ensemble at high ligand concentration.  Focusing on nCaM, we
see that helix A, the N-terminal end of helix B, and the B-C linker become more
flexible upon \ca-binding, while helix C and helix D show little change in
flexibility.  The temperature factors of the corresponding regions in cCaM show
qualitatively similar behavior.  

All four binding loops, on the other hand,
become more rigid upon \ca coordination. The difference in flexibility upon binding
is largest for loop IV due to its large fluctuations in the unligated ensemble.
Greater entropic stabilization of loop IV in the unligated state explains
its relatively small binding affinity.\cite{gifford:07}
Furthermore, accounting for differences in loop entropy 
completes the rationalization of the
binding free energies to the loops of CaM: while the value of
$\epsilon_\mathrm{o}$ is dominated by the energetic stabilization of
binding to the open state, the value of $\epsilon_\mathrm{c}$ reflects
the degree of conformational entropy of the loop in the unligated ensemble.

The flexibility of individual residues are local order parameters that
characterizes residue-specific conformational changes upon \ca
binding.\cite{tripathi:09} To qualitatively understand CaM's structural changes
along the binding curve, we compare the fluctuations of the $i^{\mathrm{th}}$
residue to a two state reference rmsf, $\beta_0(i)$, given by average
\begin{equation}\label{eq:rms avg}
  \beta_{\mathrm{0}}(i) = 
  \langle f_{\mathrm{b}} \rangle \beta_{\mathrm{o}}(i) 
  + (1 - \langle f_{\mathrm{b}} \rangle)\beta_{\mathrm{c}}(i),
\end{equation}
where the rmsf of the open ensemble, $\beta_\mathrm{o}(i)$, and the closed
ensemble, $\beta_\mathrm{c}(i)$, are weighted by the fractional occupancy of the
binding sites $\langle f_\mathrm{b} \rangle = \langle n_\mathrm{b}\rangle/2$.
The structural ordering of a residue at any concentration can be characterized
as early or late compared to mean flexibility $\beta_0(i)$ evaluated at the
corresponding value of $\langle f_\mathrm{b}\rangle$.  For example,
Fig.\ref{fig:ncam_ccam rms variation}  
shows the simulated rmsf of each residue at $K_\mathrm{d}$, as well as the
reference fluctuations evaluated at $\langle f_\mathrm{b} \rangle = 1/2$.
Although the \ca occupancy of the binding loops is only 50\%, the local
environment of helix A and the B-C linker of nCaM as well as corresponding helix
E and F-G linker of cCaM is already similar to that of the open state ensemble.
This ``early'' transition to the open ensemble is a reflection of the allosteric
cooperativity.  In contrast, the average structural order of the binding loops
is similar to the weighted average of the open and closed state flexibility.
The exception is the $\beta$-sheet in the C-terminal end of loop IV which takes
on the open state structure at higher ligand concentrations.  This ``late''
transition is in harmony with its lower binding affinity.

\color{black}
\section*{Concluding Remarks}
In this paper, we introduce a method to simulate binding curves involving a
protein that undergoes a conformational change upon binding.  This approach
allows us to identify the structural origins of binding affinity and to quantify
allosteric cooperativity within a simple coarse-grained description of the
protein dynamics.  In this implicit ligand model, the protein conformation
modulates the protein-ligand interactions through effective ligand-mediated
contacts among residues in the binding site.  
The influence of ligand
concentration on the effective binding strength is described through 
its uniform chemical potential.

Applying this approach to CaM, we find that this model can distinguish the
binding properties of the two domains of CaM: binding loops I and II of nCaM
have similar affinities, while in cCaM, binding loop III has significantly
greater affinity than loop IV.  The broader range of binding affinities in cCaM
accounts for its greater cooperativity.  Simulated populations of the ligation
states as a function of concentration are accurately described by the MWC model
with appropriate binding free energies for the individual loops.  These binding
free energies are average properties of the simulated ensemble and are not
obvious solely from the open and closed structures.  While the simulated binding
thermodynamics is well-described by the MWC model, this simple analysis can
obscure complexities in the free energy landscape.  In separate publication, we
describe how subtle differences in the topology and stability of the two domains
lead to distinct simulated mechanisms for \ca-free domain opening for nCaM and
cCaM (submitted). In particular, we find that cCaM unfolds more readily than
nCaM during the domain opening transition under similar conditions,  
\color{black}%
a result consistent with the lower thermal stability of the C-terminal domain
in the intact protein.\cite{masino:00,rabl:02}
\color{black}
Although
the unfolded conformations play a minor role in the binding thermodynamics
described in this paper (aside from modifying the binding free energies to the
open and closed states), global folding and unfolding in the domain opening
transition likely has a significant qualitative influence on the binding
kinetics.
This is a problem we plan to address in future work.


\begin{acknowledgments}
  We would like to thank Daniel Gavazzi for help in figure preparation. 
  Financial support from the National Science Foundation Grant No. MCB-0951039 is gratefully
  acknowledged.
\end{acknowledgments}

%

\pagebreak
\widetext
\setcounter{equation}{0}
\setcounter{figure}{0}
\setcounter{table}{0}
\makeatletter


\textbf{\large Supporting Information:
 Coarse-grained molecular simulations of allosteric cooperativity}






\section*{One-dimensional simulated free energy}

\begin{figure}[h]
  \renewcommand{\figurename}{Figure S\!\!}
  \begin{center}
    \includegraphics[]{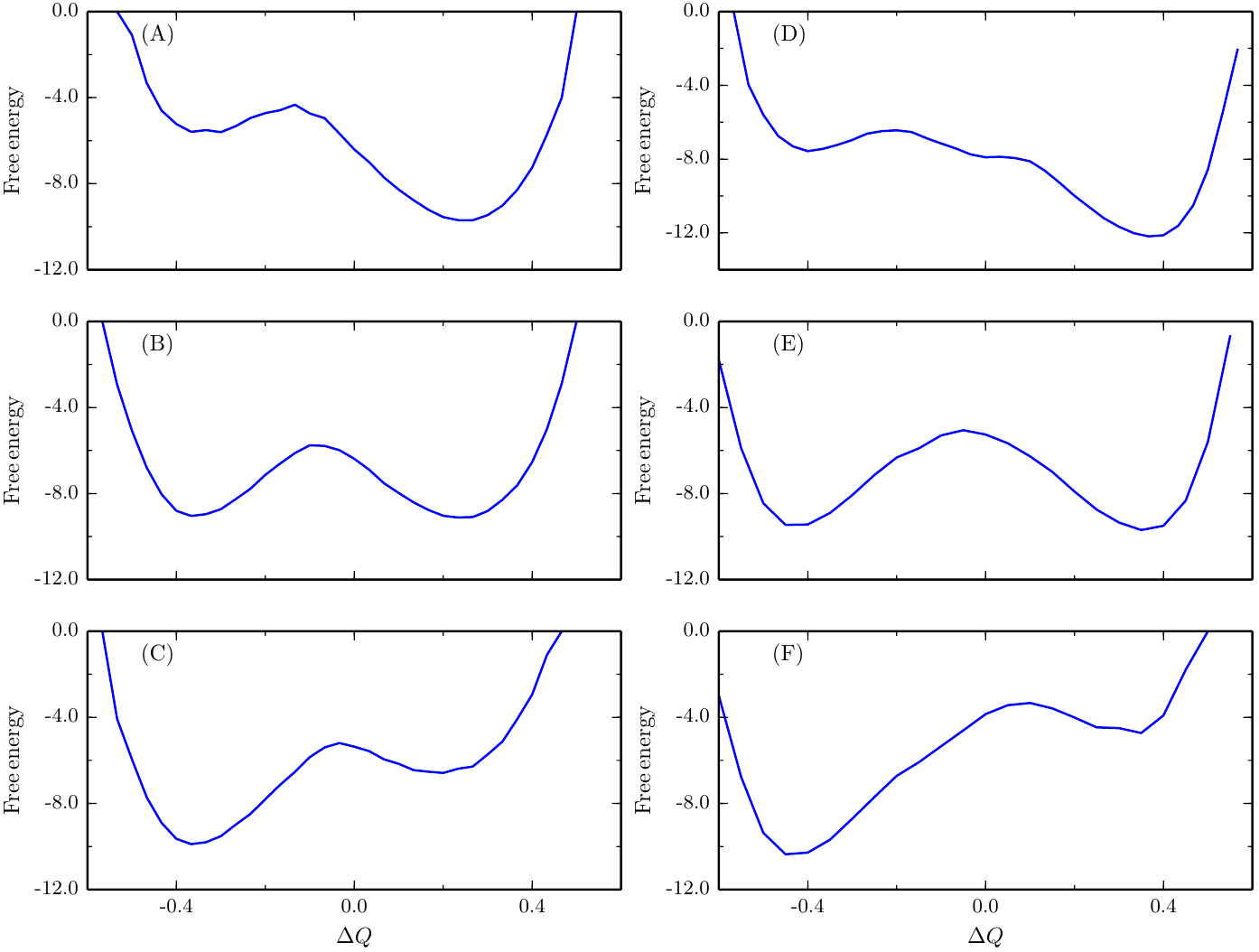}
    \caption{ 
      Simulated free energy for nCaM (A,B,C) and cCaM (D,E,F)
      corresponding to the ensemble of unligated (top), singly ligated (middle)
      and fully saturated (bottom) conformations.
      The x-axis represents simulated progress
      coordinate $\Delta Q = Q_\mathrm{{closed}} - Q_\mathrm{{open}}$ and the 
      y-axis represents simulated free energy in units of ${k_\mathrm{{B}}T}$. }
      \label{fig:ncam_ccam_1d_free_en}
  \end{center}
\end{figure}

Simulated free energy in terms of one-dimensional progress coordinate, 
$\Delta Q = Q_\mathrm{closed} - Q_\mathrm{open}$,
as shown in Fig.S1, illustrates that for both domains, the closed
state is more stable in the unligated ensemble. 
Binding of first ligand stabilizes both the closed and open states but 
the high affinity open state is stabilized to a greater extent
due to its structural compatibility with the ligand.
In the fully saturated ensemble, the open state has greater stability.

\section*{Exploring ligand contact strength and range}

For the results presented in the paper, we made specific
choices for the ligand-mediated contact strength, $\mathrm{c_{lig}} = 2.5$,
and interaction range, $\sigma_{ij} = (0.1) r_{ij}^0$, respectively. 
As shown in Fig.S2, with the increase of $\mathrm{c_{lig}}$ and $\sigma_{ij}$,
the value of $K_\mathrm{d}$ for individual loops decreases.
However, the slope of the binding transition curve at a
concentration for which $p_\mathrm{bound} = 0.5$ remains the same.
Here, we only show results for binding loop I as an illustration. 

\begin{figure}[h]
  \renewcommand{\figurename}{Figure S\!\!}
  \begin{center}
    \includegraphics[]{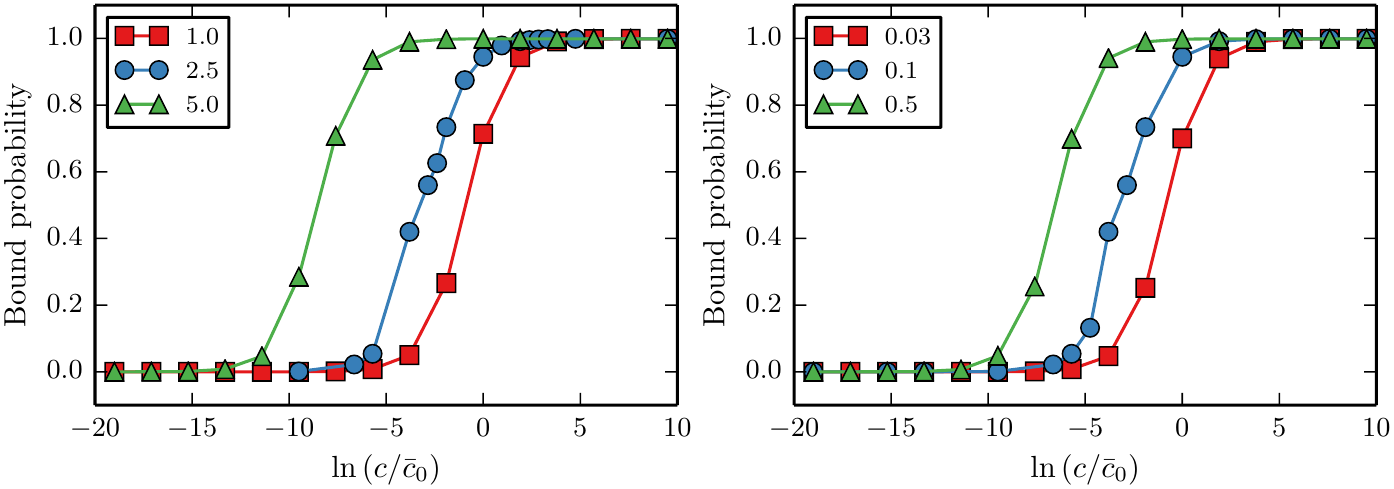}
    \caption{
      Simulated binding curves for loop I of nCaM with
      varying $\mathrm{c_{lig}}$ (left) and
      $\sigma_{ij}$ (right). 
      Consistent behavior is
      observed for other binding loops (data not shown).}
    \label{fig:p_bound_with_c_lig_sigma}
  \end{center}
\end{figure}

\end{document}